\documentclass[twocolumn]{aastex631}

\usepackage{color}

\shorttitle{Radiative CSM Acceleration in Interacting SNe}
\shortauthors{Tsuna et al.}

\begin{document}

\title{Radiative Acceleration of Dense Circumstellar Material in Interacting Supernovae}

\correspondingauthor{Daichi Tsuna}
\email{tsuna@caltech.edu}

\author[0000-0002-6347-3089]{Daichi Tsuna}
\affiliation{TAPIR, Mailcode 350-17, California Institute of Technology, Pasadena, CA 91125, USA}
\affiliation{Research Center for the Early Universe (RESCEU), School of Science, The University of Tokyo, 7-3-1 Hongo, Bunkyo-ku, Tokyo 113-0033, Japan}

\author[0000-0002-5358-5642]{Kohta Murase}
\affiliation{Department of Physics, Pennsylvania State University, University Park, Pennsylvania 16802, USA}
\affiliation{Center for Multimessenger Astrophysics, Institute for Gravitation and the Cosmos, Pennsylvania State University, University Park, Pennsylvania 16802, USA}
\affiliation{Department of Astronomy and Astrophysics, Pennsylvania State University, University Park, Pennsylvania 16802, USA}
\affiliation{School of Natural Sciences, Institute for Advanced Study, Princeton, NJ 08540, USA}
\affiliation{Center for Gravitational Physics and Quantum Information, Yukawa Institute for Theoretical Physics, Kyoto, Kyoto 16802, Japan}

\author[0000-0003-1169-1954]{Takashi J. Moriya}
\affiliation{National Astronomical Observatory of Japan, 2-21-1 Osawa, Mitaka, Tokyo 181-8588, Japan}
\affiliation{School of Physics and Astronomy, Faculty of Science, Monash University, Clayton, Victoria 3800, Australia}

\begin{abstract}
Early-time light curves/spectra of some hydrogen-rich supernovae (SNe) give firm evidence on the existence of confined, dense circumstellar matter (CSM) surrounding dying massive stars. We numerically and analytically study radiative acceleration of CSM in such systems, where the radiation is mainly powered by the interaction between the SN ejecta and the CSM. We find that the acceleration of the unshocked dense CSM ahead of the shock is larger for massive and compact CSM, with velocities reaching up to $\sim 10^3\ {\rm km\ s^{-1}}$ for a CSM of order $0.1\ M_\odot$ confined within $\sim 10^{15}$ cm. We show that the dependence of the acceleration on the CSM density helps us explain the diversity of the CSM velocity inferred from the early spectra of some Type II SNe. For explosions in even denser CSM, radiative acceleration can affect the dissipation of strong collisionless shocks formed after the shock breakout, which would affect early non-thermal emission expected from particle acceleration.
\end{abstract}


\section{Introduction}
Recent observations of supernovae (SNe) have deepened our understanding of the dramatic end stages of stellar evolution. A good fraction of hydrogen-rich (Type II) SNe show signatures of dense circumstellar matter (CSM), such as the narrow emission lines \citep{Schlegel90,Filippenko17,Pastorello08,Khazov16,Yaron17,Bruch21,Bruch22}, and light curves at early times \citep[e.g.,][]{Morozova17,Morozova18,Das17,Forster18, Moriya18}. Outbursts are also observed in some Type II SNe months to years before explosion, which may be related to the origin of the CSM \citep{Fraser13,Ofek14,Strotjohann21,Jacobson22,Matsumoto22,Tsuna23}.

The appearance of these SNe is heterogeneous in duration, luminosity and spectral features \citep[e.g.,][]{Taddia13,Smith17}. For example, among Type IIn SNe there are long-lasting ones like SN 1988Z and SN 2005ip \citep{Stathakis91,Turatto93,Stritzinger12,Smith17_2005ip}, transitional ones like SN 1998S \citep{Leonard00,Fassia01}, ones accompanying a plateau light curve like SN 1994W \citep{Sollerman98}, and superluminous ones like SN 2010jl \citep{Stoll11,Frannson14,Ofek14_10jl}. The CSM of the more common Type II SNe is also diverse, with mass-loss rates ranging from a few $10^{-4}$ up to $1\ M_\odot\ {\rm yr}^{-1}$ \citep{Yaron17,Boian20}. These indicate a large spread of the progenitor's mass-loss history, with a vast range of masses released at a broad timescale from months to centuries before core-collapse.

The extreme CSM densities of these SNe lead to large dissipation of the kinetic energy of the SN ejecta via shocks. 
It is expected that the shock has a transtion from radiation-mediated to collisionless, which may be accompanied by the onset of particle acceleration and resulting non-thermal emission \citep{Murase11,Katz12}. 
Following shock breakout the outward photon pressure, from shock breakout and later CSM interaction, can lead to the bulk acceleration of the ambient CSM to large velocities \citep{Katz12}. The bulk acceleration of the CSM may also cause a negative feedback to the dissipation via collisionless shocks or even delay the sub-shock formation itself \citep{Murase19}.

Similar to the mass-loss history, the CSM velocity in interacting SNe is also found to be diverse. Early compilations of Type IIn SNe found CSM velocities in the broad range of $100$--$1500$ km s$^{-1}$ \citep{Kiewe12,Taddia13}. More recently, \cite{Boian20} conducted spectral fitting for the dense CSM identified by early flash spectroscopy of Type II SNe, using the non-LTE radiative-transfer code CMFGEN \citep{Hillier98,Hillier12}. The inferred velocities of the dense CSM also vary from $<100\ {\rm km \ s^{-1}}$ to $800\ {\rm km \ s^{-1}}$, in many cases much higher than seen in winds of red supergiants (RSGs). Radiative acceleration may be a link between the densities and velocities of observed interacting SNe. 

By numerical and analytical modelling, we quantitatively study the acceleration for different cases of mass and extent of the CSM. We find that this process can reproduce the diverse CSM velocity observed in spectra of young Type II SNe. We also find that CSM acceleration can greatly suppress the shock dissipation at early phases, which can have important consequences for early non-thermal emission due to CSM interaction.

This work is constructed as follows. In Section \ref{sec:numerical} we first investigate this problem numerically, using the radiation hydrodynmics code SNEC (SuperNova Explosion Code; \citealt{Morozova15}). Using the simulation results, in Section \ref{sec:analytical} we construct an analytical model of radiative acceleration, and compare our results with CSM velocities obtained from spectra of interacting SNe. In Section \ref{sec:nonthermal} we discuss the effect of CSM acceleration on the onset and dissipation of collisionless shocks. We conclude in Section \ref{sec:conclusion}.

\section{Numerical Study}
\label{sec:numerical}
\begin{table*}
\centering
\begin{tabular}{c||ccc|ccccc}
& CHIPS & & & SNEC & \\ \hline
Model parameters & $r_*$ [cm] & $r_{\rm CSM}$ [cm] & $M_{\rm CSM}$ [${\rm M}_\odot$] & $t_{\rm bo}$ [day] & $t_{\rm ph}$ [day] & $E_{\rm rad}$ [erg] & $\langle v_{\rm sh}\rangle$ [km s$^{-1}$] & $\langle f_{\rm rad}\rangle$ \\ \hline
$t_{\rm erup}=-10$ yr& $7.1\times 10^{13}$& $1.6\times10^{15}$ & $0.063$& $1.9$ & 21 & $1.8\times 10^{49}$ & 5800 & 0.85 \\
$t_{\rm erup}=-5$ yr& $5.5\times 10^{13}$& $8.0\times10^{14}$ & $0.098$& 2.4 &17 & $2.5\times 10^{49} $  & 5400 & 0.87\\
$t_{\rm erup}=-3$ yr& $5.9\times 10^{13}$& $4.8\times10^{14}$ &$0.15$& 4.6 &13 &$3.2\times 10^{49} $ & 5200 & 0.80\\
$t_{\rm erup}=-1$ yr& $5.1\times 10^{13}$& $1.7\times10^{14}$ &$0.5$& 4.2 & 6.4 & $2.1\times 10^{49}$ & 5500 & $0.15$
\end{tabular}
\caption{Model parameters simulated by CHIPS and SNEC in this work. Columns are border between star and CSM, break radius of the double power-law density profile, total CSM mass outside $r_*$, time of shock breakout, time when photosphere enters the shock downstream, energy of radiation emitted up to $t_{\rm ph}$, mean shock velocity and radiation conversion efficiency (see equations \ref{eq:mean_rad_eff} and \ref{eq:mean_vel} in Section \ref{sec:analytical}).}
\label{table:NumericalParameters}
\end{table*}

In this section we first outline the procedures for simulating the CSM originating from mass eruption of the progenitor by CHIPS, and the subsequent SN explosion by SNEC. We then look into the results of the SNEC simulations, focusing on the radiation output due to CSM interaction and the resulting acceleration of the CSM.
\subsection{Generating the CSM}
The CSM, and especially its density profile, should depend on the formation mechanism. Here we adopt a CSM model generated by mass eruption due to sudden energy injection in the envelope. We simulate the eruption using the one-dimensional radiation hydrodynamics code in CHIPS \citep[][for details see also \citealt{Kuriyama20}]{Takei22}.

CHIPS solves the mass eruption of a star specifically triggered by energy injection at the base of an arbitrarily computational region. The initial profile is given from a one-dimensional stellar model in hydrostatic equilibrium, which is interpolated and mapped into a grid of 10000 meshes. Internal energy is injected into the base of the computational region, and its hydrodynamical response is solved by the radiation hydrodynamical code in Lagrangian coordinates. After energy injection a pulse forms and propagates outwards, steepening into a shock. Once the shock reaches near the stellar surface it breaks out, ejecting the outer part of the envelope. The discontinuity at the shock is smoothed by artificial viscosity in the code, as done in most hydrodynamical simulations. Gas and radiation are assumed to always be in thermal equilibrium, whose temperature is obtained from the internal energy and density with the HELMHOLTZ equation of state \citep{Timmes00}. This single-temperature assumption for the erupted CSM is generally valid, except for the outermost part of the erupted material whose densities significantly deviate from the values in the star.

For the initial progenitor we adopt a RSG of initial mass $15M_\odot$ and solar metallicity, simulated up to core-collapse by MESA \citep{Paxton11,Paxton13,Paxton15,Paxton18,Paxton19}. We adopt the \verb|example_make_pre_ccsn| test suite in revision 12778, that includes mixing of \cite{Henyey65} with a mixing length parameter (\verb|mixing_length_alpha|) of 3 (1.5) for the hydrogen rich (poor) region, and a Dutch wind prescription with a scheme by \citet{deJager88} for cooler effective temperatures  of $<10^4$ K. We adopt the progenitor that is evolved until core-collapse as initial condition for the eruption simulation, but we verified that the structure of the envelope does not change much within the last decades of its life. This progenitor has a radius of $670\ R_\odot$ and total mass of $12.8\ M_\odot$, with $4.9$ and $7.9\ M_\odot$ occupied by the helium core and hydrogen-rich envelope respectively.

It is plausible to assume that the energy triggering envelope eruption is sourced from the stellar core that is undergoing nuclear burning \citep[e.g.][]{Quataert12,Smith_Arnett_14,Shiode14,Fuller17,Fuller18,Wu21,Leung21}. We thus extract the hydrogen-rich envelope as the computational region, and as a representative case inject $1.4\times 10^{47}$ erg ($30$\% of the envelope's binding energy) at the inner boundary at a constant rate over 1000 seconds. We follow the hydrodynamical response of the envelope, and record the profiles at $1,\ 3,\ 5$ and $10$ years after energy injection. We label these models as $t_{\rm erup}=-1,\ -3,\ -5,\ -10$ yr, where $t_{\rm erup}$ is the time of energy injection relative to the time of explosion.

\begin{figure}
\centering
\includegraphics[width=\linewidth]{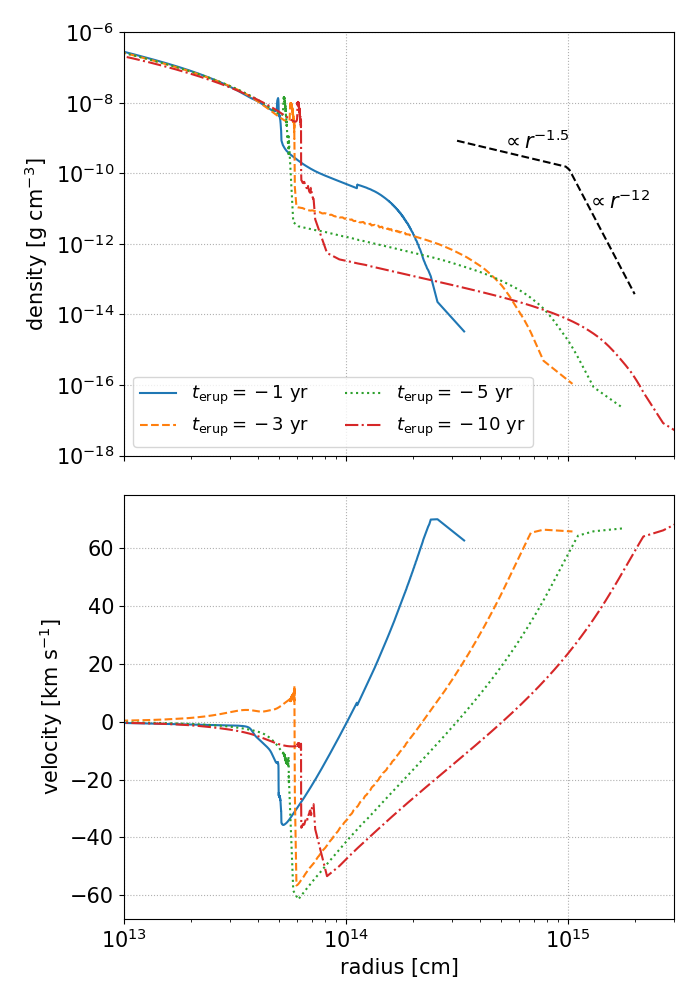}
\caption{Density and velocity profiles of the CSM models considered in the numerical study. The discontinuities in the profiles around $5$--$7\times 10^{13}$ cm (see Table \ref{table:NumericalParameters}) are regarded as the boundaries between the star and the CSM, $r_*$. Matter near the stellar surface oscillates on a dynamical timescale, which changes $r_*$ and velocity near $r_*$ for different $t_{\rm erup}$.}
 \label{fig:rho_CSM}
 \end{figure}

We show the density profiles of these CSM in Figure \ref{fig:rho_CSM}. The profile is generally well fitted by a double power-law, with an inner shallow profile of $\rho\propto r^{-1.5}$ rolling over to a steep profile of about $\rho\propto r^{-12}$ \citep{Tsuna21,Ko22}. An exception is the $t_{\rm erup}=-1$ yr model, where the CSM has not experienced fallback that characterizes the $\rho\propto r^{-1.5}$ profile. Thus we conduct least-squares fits of the profile as in \cite{Tsuna21}, but adding the inner power-law index as a fitting parameter. The border between the star and the CSM $r_*$ (inner edge of each fit) is set to the radii given in Table \ref{table:NumericalParameters}, where the density and velocity profiles discontinuously change. The break radius $r_{\rm CSM}$ and total mass $M_{\rm CSM}$ of the CSM obtained from these fits are shown in Table \ref{table:NumericalParameters}. The CSM at around the break radius expands homologously, with a velocity of $v_{\rm CSM}\approx 50$ km s$^{-1}$ at $r=r_{\rm CSM}$. The total CSM mass decreases over time, due to the fallback of the inner bound part of the CSM. These two effects produce a broad range in the CSM density, which lead to different consequences for radiative acceleration as shown later.

\subsection{Simulating Explosions by SNEC}
\label{sec:numerical_SNEC}
To simulate the following explosion of the progenitor by SNEC, we then stitch each envelope model to the helium core in the MESA model. The eruption simulation is Lagrangian, so we use the conserved mass coordinates to replace the hydrodynamical profiles of the MESA model to those from our simulations. The Kelvin-Helmholtz timescale of this RSG progenitor is $\approx 80$ years, so the hydrodynamical profiles of the envelope models have not settled down to the initial phase yet. The density profiles passed over to SNEC are thus not smooth at the helium core, containing discontinuities by typically a factor of $\approx 2$. However, the dynamics of the CSM interaction should be governed only by the outer layers of the star (ejecta) where the reverse shock formed by the interaction propagates, and not by the detailed profile at the inner regions.

We excise the innermost region of mass $1.8\ M_\odot$, slightly outside the silicon-oxygen interface as suggested by \cite{Morozova18}. We then inject an internal energy of $10^{51}$ erg over $0.1$ seconds in the inner $0.1\ M_\odot$ as a thermal bomb. The computational regions for the models of $t_{\rm erup}=-3,-5$ yr are meshed with 3000 grids, and the models of $t_{\rm erup}=-1,-10$ yr with a coarser 1000 grids.

\begin{figure}
 \centering
 \includegraphics[width=\linewidth]{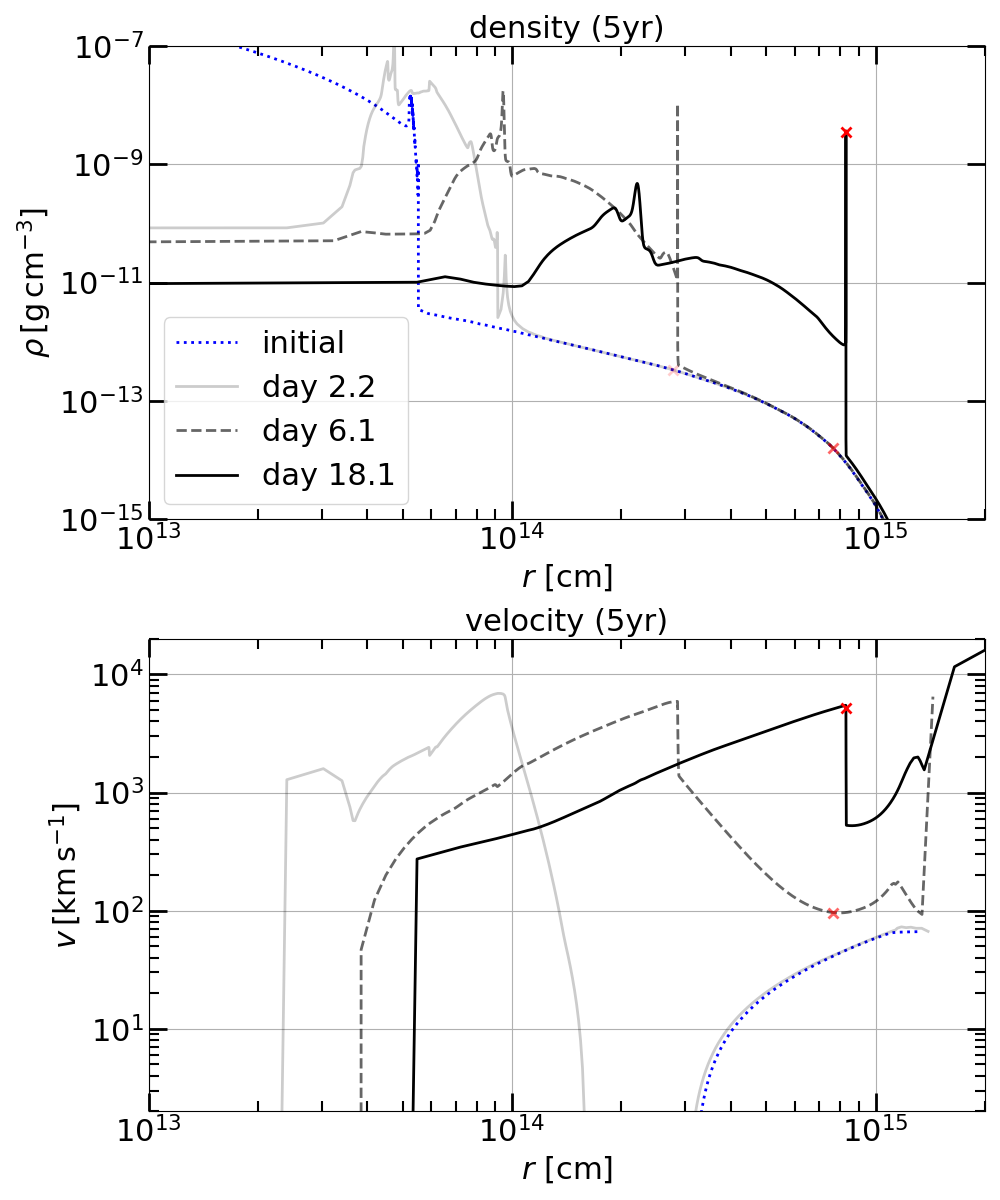}
\caption{Density and velocity profiles simulated by SNEC, for the CSM model with $t_{\rm erup}=-5$ yr. The blue dotted lines show the initial profile at core-collapse, and the red crosses show the location of the photosphere.}
 \label{fig:SNECprofile_5yr}
 \end{figure}
 
\begin{figure*}
   \centering
    \begin{tabular}{cc}
     \begin{minipage}[t]{0.5\hsize}
    \centering
 \includegraphics[width=\linewidth]{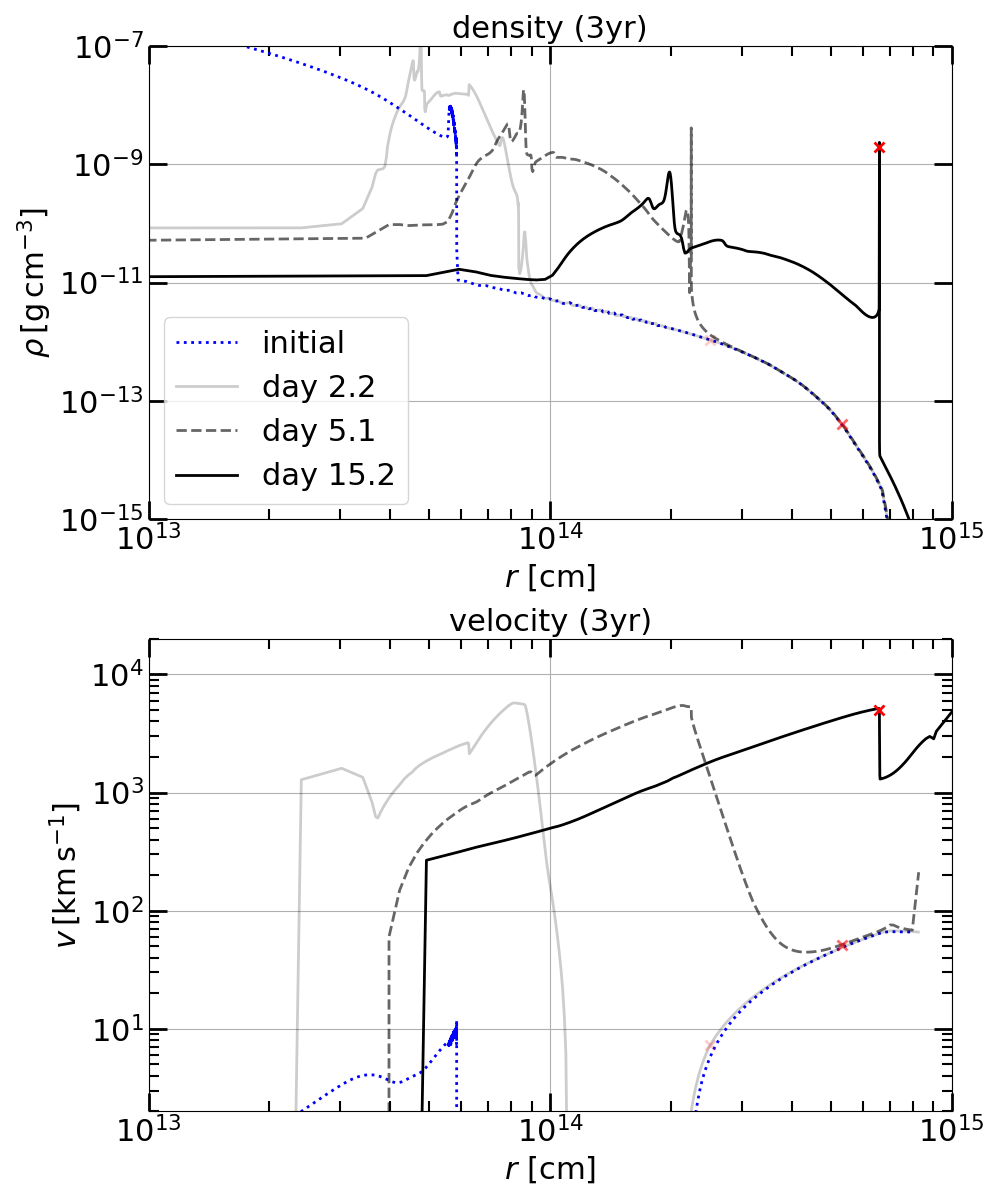}
    \end{minipage}
     \begin{minipage}[t]{0.5\hsize}
    \centering
 \includegraphics[width=\linewidth]{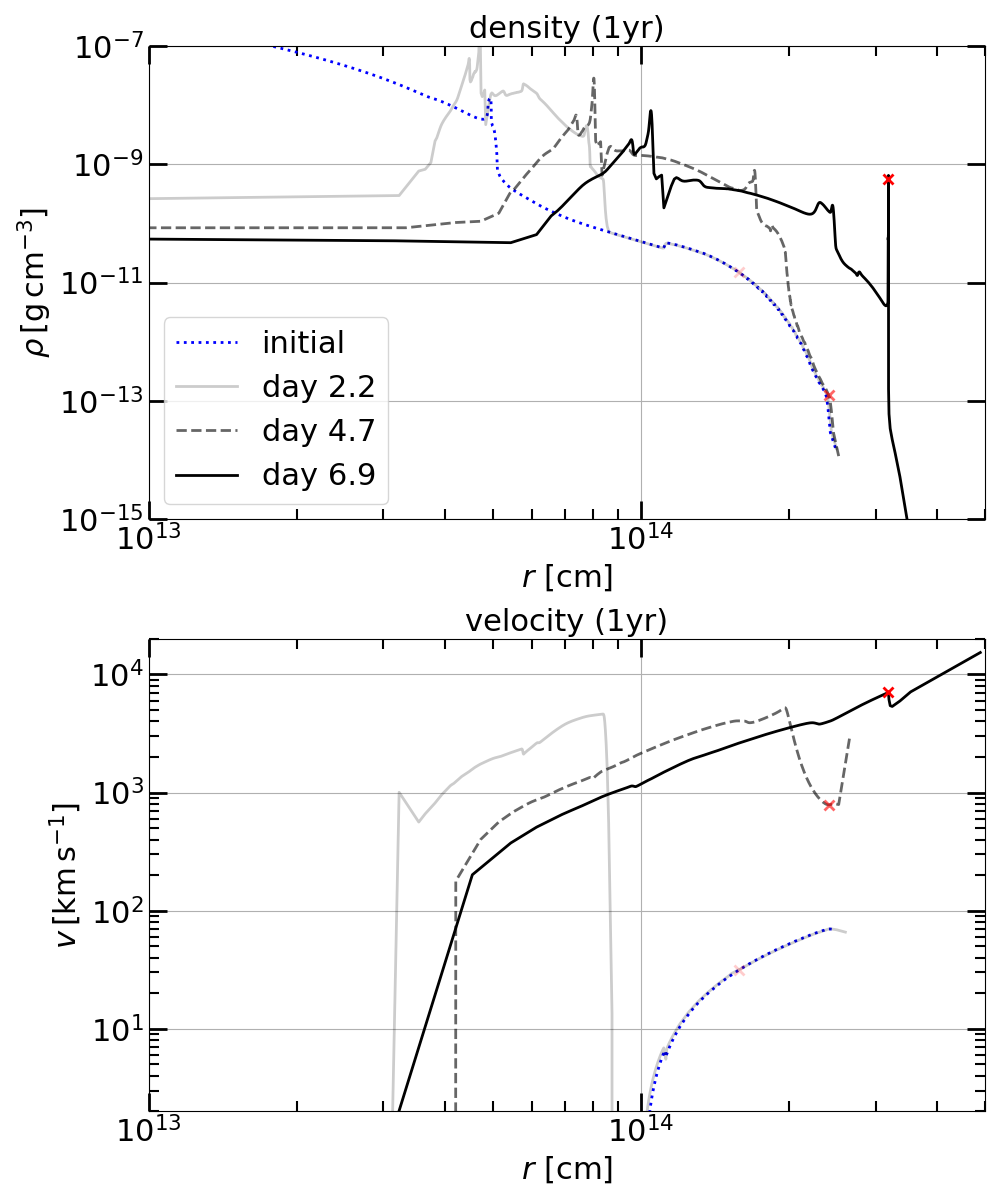}
    \end{minipage}
    \end{tabular}
\caption{Same as Figure \ref{fig:SNECprofile_5yr}, but for the CSM models with $t_{\rm erup}=-3$ yr and $-1$ yr. Note that we plot the profiles for different epochs from Figure \ref{fig:SNECprofile_5yr}.}
 \label{fig:SNECprofile_3_1yr}
 \end{figure*}
 
We show the evolution of the density and velocity profiles in Figures \ref{fig:SNECprofile_5yr} and \ref{fig:SNECprofile_3_1yr} for the models for $t_{\rm erup}=-5, -3, -1$ yrs. For all models the shock formed in the center crosses the star at $\approx 1.5$ days from core-collapse. For the $t_{\rm erup}=-5$ yr model, the CSM is not so dense and radiation starts to escape shortly after, from about day $2$. The inner CSM near the shock is then accelerated, but the velocity and density profiles at the far upstream are unchanged from the initial ones in blue dotted lines. A shock emerges a few days after breakout, which can be seen from the velocity jump on day 6.1 at $r\approx 3\times 10^{14}$ cm. A thin shell of radiatively cooled shocked material can be seen as a spike in the density profile at the shock downstream. We show the photosphere recorded by SNEC at optical depth $\tau=2/3$ as red crosses\footnote{In SNEC the opacity is the Rosseland mean, derived from the OPAL opacity table \citep{Iglesias96} at high temperatures (log $T>3.75$), and a model of \cite{Ferguson05} at low temperatures (for details see \citealt{Morozova15}).}. The photosphere near the outer edge of the CSM would eventually enter the shock downstream, at day 17 for this model. We define this time as $t_{\rm ph}$, and present its values for the CSM models in Table \ref{table:NumericalParameters}.

For the $t_{\rm erup}=-3$ yr model, a similar behaviour is seen as the $t_{\rm erup}=-5$ yr model, except that the density profile of the unshocked CSM is slightly modified from the initial one at late phases due to the stronger acceleration. For the model of $t_{\rm erup}=-1$ yr with most compact CSM, the behaviour is quite different. The much denser CSM prolongs the shock breakout until around day $4$, when the shock has travelled through most of the CSM out to a radius of $r_{\rm sh}\approx 1.7\times 10^{14}\ {\rm cm}\ (v_{\rm sh}/5000\ {\rm km\ s^{-1}})(t/4\ {\rm day})$, where $v_{\rm sh}$ is the shock velocity. Just after breakout, the unshocked CSM rapidly accelerates to velocity comparable to the shock velocity. The density profile has drastically changed from the initial one. The shock eventually catches up with the accelerated photosphere at around day 6.

 \begin{figure}
 \centering
 \includegraphics[width=\linewidth]{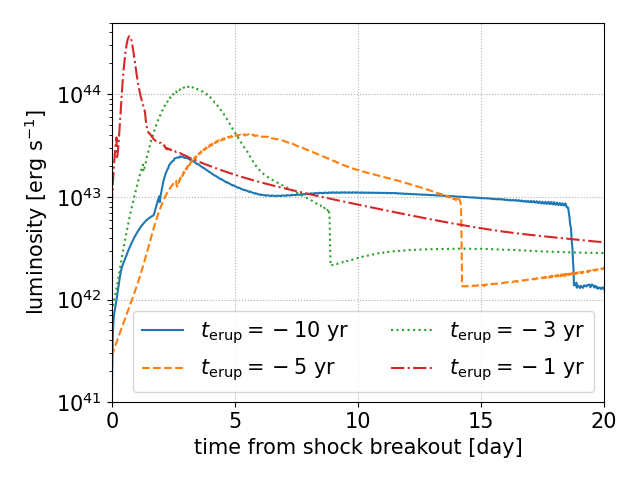}
\caption{Light curves of the four models calculated by SNEC. The breaks in $t_{\rm erup}=-3, -5, -10$ yr models are when the shock sweeps the photosphere in the CSM.}
 \label{fig:L_bol}
 \end{figure}
 
The interaction between the CSM efficiently converts the kinetic energy into radiation, as observed in the light curves of Figure \ref{fig:L_bol}. The radiated energy during the interaction phase is in the range of $(1.8$--$3.2)\times 10^{49}$ erg s$^{-1}$. The shock eventually sweeps the dense part of the CSM at $t\approx t_{\rm ph}$, and the interaction terminates soon after. For $t_{\rm erup}=-3,-5,-10$ yrs this appears as a steep drop in the light curve, after which the light curve is powered by recombination in the envelope as in Type II-P SNe. For the $t_{\rm erup}=-1$ yr model, the shocked region is still optically thick and the light curve is prolonged by the cooling emission of this region.

We note that SNEC assumes that the shock-heated gas radiatively cools much faster than adiabatically. This is valid if the cooling timescale of downstream gas is shorter than the dynamical time. \cite{Chevalier12} finds that for a wind profile CSM ($\rho\propto r^{-2}$), this is satisfied for mass-loss rates of
\begin{eqnarray}
\dot{M}&\gtrsim& 5\times 10^{-4}\ M_\odot\ {\rm yr}^{-1} \left(\frac{v_{\rm sh}}{5000\ {\rm km\ s^{-1}}}\right)^3 \nonumber \\
&&\times \left(\frac{t}{10\ {\rm day}}\right) \left(\frac{v_{\rm CSM}}{50\ {\rm km\ s^{-1}}}\right)    \label{eq:fast_cooling}
\end{eqnarray} 
for a shock velocity of $v_{\rm sh}\lesssim 10^4$ km s$^{-1}$. Although our CSM models slightly differ from a wind profile, this condition is always satisfied in our models with $M_{\rm CSM}/(-t_{\rm erup})\approx (6$--$500)\times 10^{-3}\ M_\odot\ {\rm yr}^{-1}$. Thus the simulations accurately reflect the actual contribution of radiation produced in the shock downstream.

\subsection{Acceleration of the CSM}
\label{sec:numerical_acc}

 \begin{figure}
 \centering
 \includegraphics[width=\linewidth]{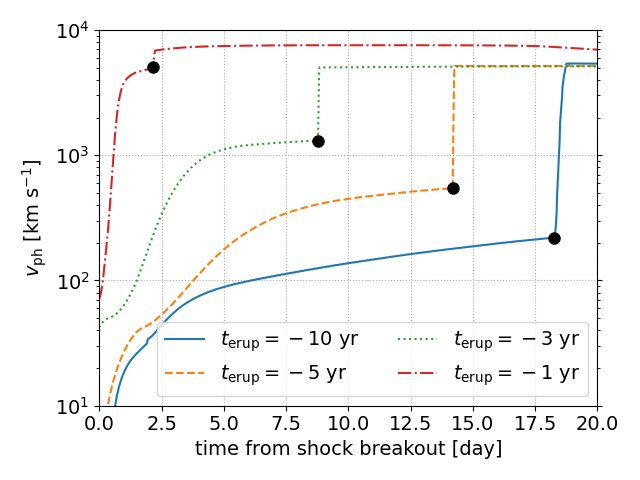}
\caption{Time evolution of the photospheric velocity $v_{\rm ph}$ of our SNEC models. CSM at the photosphere experiences radiative acceleration before the vertical jump, when the photosphere enters the shock downstream at the epochs shown as black dots.}
 \label{fig:v_ph_SNEC}
 \end{figure}

To study the radiative acceleration, we measure the velocity of the CSM at the photosphere $v_{\rm ph}$. This is because we expect the observed spectral properties, including the line width, to be characterized by the material around the photosphere.

The temporal evolution of $v_{\rm ph}$ is plotted in Figure \ref{fig:v_ph_SNEC}. For all cases the CSM at the photosphere smoothly accelerates until the epoch shown as black dots in the figure, after which its velocity steeply jumps as it is is engulfed by the shock. The degree of acceleration is significantly altered when $t_{\rm erup}$ is changed. The long $t_{\rm erup}=-10$ yr model, with the photosphere farther away and relatively small CSM mass, results in a weak, gradual acceleration to at most $\approx 200$ km s$^{-1}$, while a model of compact CSM with $t_{\rm erup}=-3$ yr results in a large acceleration of $\gtrsim 1000$ km s$^{-1}$. The extreme case with $t_{\rm erup}=-1$ yr results in acceleration of the CSM to velocities comparable to the shock, from just after breakout (see also Figure \ref{fig:SNECprofile_3_1yr}).

We can understand the above tendency from the scaling of the acceleration on the CSM parameters. When the shock with velocity $v_{\rm sh}$ sweeps through the CSM with negligible velocity out to a radius $r_{\rm sh}$, the total dissipated kinetic energy by the CSM is $\approx M_{\rm CSM}(<r_{\rm sh})v_{\rm sh}^2/2$, where $M_{\rm CSM}(<r)$ is the CSM mass enclosed within $r$. If a fixed fraction of this dissipated energy is converted to radiation and escape outwards, the energy of radiation $E_{\rm rad}$ that contributes to acceleration is $E_{\rm rad} \propto M_{\rm CSM}(<r_{\rm sh}) v_{\rm sh}^2/2$.
The total outward velocity that the CSM at radius $r$ can receive due to radiative acceleration is then $\kappa E_{\rm rad}/4\pi r^2 c$, where $\kappa$ is the opacity and $c$ is the speed of light. A smaller $-t_{\rm erup}$ (i.e., shorter interval between eruption and explosion) results in a compact (small photospheric radius) and large $M_{\rm CSM}$, which both contribute to enhancing the acceleration due to the dependence $\propto M_{\rm CSM}(<r_{\rm sh})r^{-2}$. In the next section, we study this acceleration more quantitatively by constructing an analytical model.

\section{Analytical Model}
\label{sec:analytical}
In this section we construct an analytical model of radiative acceleration, with a more simplified setting to realize a more efficient exploration of the parameter space of the CSM. We first introduce the model setup, and discuss some of the approximations in the model using the results we obtained from the numerical study.

We parameterize the CSM with its total mass $M_{\rm CSM}$ and extent $R_{\rm CSM}=(-t_{\rm erup})v_{\rm CSM,0}$, where we adopt $v_{\rm CSM,0}=50$ km s$^{-1}$ for the CSM expansion velocity before acceleration. This value is motivated from the velocity of near the outer edge of the CSM simulated in Section \ref{sec:numerical}, and is comparable to the surface escape velocity of RSGs. We consider a CSM profile of a power-law with index $s$,
\begin{eqnarray}
    \rho_{\rm CSM}(r) = Dr^{-s}\  (R_*<r<R_{\rm CSM}).
\end{eqnarray}
where the normalization $D$ is given from $M_{\rm CSM}, R_{\rm CSM},$ and $R_*$ as
\begin{eqnarray}
     D = \frac{(3-s)M_{\rm CSM}}{4\pi(R_{\rm CSM}^{3-s}-R_*^{3-s})}.
\end{eqnarray}

After the core-collapse at $t=0$, a radiation-mediated shock travels outwards, with velocity $v_{\rm sh}$, through a star surrounded by the dense CSM. The shock eventually breaks out either in the star or the CSM depending on the optical depth of the (unshocked) CSM $\tau_{\rm CSM} = \int_{R_*}^{R_{\rm CSM}} \kappa \rho_{\rm CSM}(r) dr$, where $\kappa$ is the opacity. Approximating $\kappa$ as a constant of radius, the breakout radius $r_{\rm bo}$ where $\tau(r_{\rm bo})=c/v_{\rm sh}$ is
\begin{eqnarray}
    r_{\rm bo} \approx {\rm max}\left[R_*, \left(R_{\rm CSM}^{1-s}+\frac{s-1}{\kappa D}\frac{c}{v_{\rm sh}}\right)^{1/(1-s)}\right],
\end{eqnarray}
where $c$ is the speed of light\footnote{We note that the optical depth for the breakout criterion should be measured from where $\tau\sim 1$, instead of $r=R_{\rm CSM}$ where $\tau=0$ \citep{Moriya12}. However, the difference of $r_{\rm bo}$ would be small for non-relativistic shocks since $c/v_{\rm sh}\gg 1$.}. The former in the square brackets is when $\tau_{\rm CSM}<c/v_{\rm sh}$, in which the breakout happens near the stellar surface. Comparison of $\tau_{\rm CSM}$ and $c/v_{\rm sh}$ at $r=R_*$ determines whether the shock breaks out near the stellar surface ($r_{\rm bo}\approx R_*$, where $r_{\rm bo}$ is the breakout radius) or in the CSM ($R_*<r_{\rm bo}<R_{\rm CSM}$).

After the shock breakout, photons escape outwards and accelerates even the unshocked CSM. We consider the photons originating from the kinetic energy of the ejecta dissipated by the CSM. When the shock is at radius $r_{\rm sh}$, the radiation energy generated up to this radius is given by~\citep{Katz12}
\begin{eqnarray}
    E_{\rm rad}(r_{\rm sh}) &=& \int_{R_*}^{r_{\rm sh}} f_{\rm rad}(r') 2\pi r'^2 \rho_{\rm CSM}(r')v_{\rm sh}(r')^2 dr' ,
\end{eqnarray}
where $f_{\rm rad}(r')$ is the radiation conversion efficiency of the shock at radius $r'$. Neglecting the light-travel time in the post-breakout phase, the radiative acceleration of the unshocked CSM at radius $r$ ($r_{\rm sh}<r \leq r_{\rm CSM}$) is
\begin{eqnarray}
    \Delta v_{\rm CSM}(r,t) &\approx& \frac{\kappa E_{\rm rad}(r_{\rm sh}(t))}{4\pi r^2 c} \label{eq:Deltav_CSM_rad}\\
    &\sim& 90\ {\rm km\ s^{-1}}\left(\frac{\kappa}{0.34\ {\rm cm^2\ g^{-1}}}\right)\nonumber \\
     &\times&\left(\frac{r_{\rm sh}}{10^{15}\ {\rm cm}}\right)^{-2}\left(\frac{E_{\rm rad}(r_{\rm sh})(t)}{10^{49}\ {\rm erg}}\right).
\end{eqnarray}
Hereafter, we employ some simplifying approximations for our model. We first assume that the shock velocity $v_{\rm sh}$ is a constant of $r$. This is appropriate when the reverse shock is sweeping the outer steep part of the SN ejecta ($\rho\propto v^{-n}$ with $n=10$--$12$ for typical SN parameters), as $v_{\rm sh} \propto t^{-(3-s)/(n-s)}$ \citep{Chevalier82} very weakly depends on time as it propagates through the CSM. This nearly constant $v_{\rm sh}$ is indeed seen in our shock capture analysis done in Section \ref{sec:nonthermal}. 

The radiation conversion efficiency $f_{\rm rad}(r)$ is a more uncertain parameter in the model. Here we assume for simplicity that this efficiency does not greatly evolve with shock radius, and adopt a fixed $\langle f_{\rm rad}\rangle$ over the entire shock propagation in the CSM. In order to infer its value and dependence on the CSM parameters, we evaluate this from the results of the four numerical simulations in Section \ref{sec:numerical_SNEC} by
\begin{eqnarray}
    \langle f_{\rm rad}\rangle \equiv \frac{\int_{t_*}^{t_{\rm ph}}Ldt}{0.5M_{\rm CSM}\langle v_{\rm sh}\rangle^2}, \label{eq:mean_rad_eff}
\end{eqnarray}
where $L$ is the luminosity and
\begin{eqnarray}
    \langle v_{\rm sh}\rangle \equiv \frac{r_{\rm sh}(t=t_{\rm ph})-R_{\rm *}}{t_{\rm ph}-t_{\rm *}} \label{eq:mean_vel}
\end{eqnarray}
is the average velocity of the shock throughout its propagation in the CSM up to the photosphere, which exists near the edge of the CSM in the setups considered in this work, and $t_{\rm ph}$ is the time at $r_{\rm ph}=r_{\rm sh}$. The time $t_*\equiv t(r_{\rm sh}=R_*)$ is defined as when the shock passes the stellar surface $R_*$ in the numerical model. 

In Table \ref{table:NumericalParameters} we show the values of $\langle f_{\rm rad}\rangle$ for the four numerical models, as well as the velocity $\langle v_{\rm sh}\rangle$. We find that the conversion efficiency $\langle f_{\rm rad}\rangle$ obtained from equation (\ref{eq:mean_rad_eff}) is high ($\approx 0.85$) for models with $t_{\rm erup}=-3, -5, -10$ years, but is much lower at $t_{\rm erup}=-1$ yr. For the latter compact and massive CSM $\langle f_{\rm rad}\rangle$ is low because the diffusion timescale in the shocked CSM is long, and most of the stored radiation is released after $t_{\rm ph}$. The diffusion time is governed by the timescale it takes for the photons to travel through the unshocked CSM \citep{Balberg11}, which is given as
\begin{eqnarray}
    t_{\rm diff} &\approx& \frac{\kappa M_{\rm CSM}}{4\pi R_{\rm CSM}^2}\frac{R_{\rm CSM}}{c} \label{eq:t_diff}\\
    &\simeq& 31\ {\rm day} \left(\frac{M_{\rm CSM}}{0.5M_\odot}\right)\left(\frac{R_{\rm CSM}}{1.7\times 10^{14}\ {\rm cm}}\right)^{-1},
\end{eqnarray}
where we adopted $\kappa=0.34\ {\rm cm^2\ g^{-1}}$ that is consistent with the opacity at the photosphere in our SNEC results. Thus in the case of $t_{\rm ph}<t_{\rm diff}$, the low efficiency can be explained if only a fraction $\sim t_{\rm ph}/t_{\rm diff}$ of the radiation actually contributes to acceleration of the CSM.

We thus adopt the efficiency $ \langle f_{\rm rad} \rangle$ in our model by the following equation
\begin{eqnarray}
    \langle f_{\rm rad} \rangle \equiv f_{\rm rad, 0}\times {\rm min}\left(1, \frac{t_{\rm ph}}{t_{\rm diff}}\right)
    \label{eq:f_rad_simplified}
\end{eqnarray}
where we set $f_{\rm rad, 0}=0.85$, $t_{\rm ph}\sim R_{\rm CSM}/v_{\rm sh}$ and $t_{\rm diff}$ as given in equation (\ref{eq:t_diff}).

With these approximations on $v_{\rm sh}$ and $f_{\rm rad}$, the radiated energy can be simplified as
\begin{eqnarray}
    E_{\rm rad}(r_{\rm sh}) &\approx & \frac{1}{2} \langle f_{\rm rad} \rangle M_{\rm CSM}(<r_{\rm sh}) v_{\rm sh}^2 \label{eq:E_rad}\\
    &\sim& 2\times 10^{49}\ {\rm erg}\left(\frac{\langle f_{\rm rad}\rangle}{0.85}\right) \nonumber \\
    && \times \left(\frac{M_{\rm CSM}(<r)}{0.1M_\odot}\right) \left(\frac{v_{\rm sh}}{5000 \ {\rm km\ s^{-1}}}\right)^2.
\end{eqnarray}
The velocity of the CSM is then determined from equation (\ref{eq:Deltav_CSM_rad}), and additional acceleration is expected due to the radial gradient on the accelerated velocity, $-v\partial v/\partial r$. From $\Delta v_{\rm CSM}\propto r^{-2}$, this contribution to the acceleration is given as $2(v_{\rm CSM, 0}+\Delta v_{\rm CSM})\Delta v_{\rm CSM}/r$. The contribution to the velocity is given by integrating this over time from shock breakout as
\begin{eqnarray}
    \Delta v_{\rm grad}(r,t) &\approx& \frac{2}{r}\int_{t_{\rm bo}}^t (v_{\rm CSM, 0}+\Delta v_{\rm CSM})\Delta v_{\rm CSM} dt' \nonumber \\
    &=& \frac{2\Delta v_{\rm CSM} t}{r}\left\{\frac{v_{\rm CSM, 0}}{4-s}\left[1-\left(\frac{t_{\rm bo}}{t}\right)^{4-s}\right] \right. \nonumber \\
    &&\left. + \frac{\Delta v_{\rm CSM}}{7-2s}\left[1-\left(\frac{t_{\rm bo}}{t}\right)^{7-2s}\right] \right\},
    \label{eq:Deltav_CSM_grad}
\end{eqnarray}
where $t_{\rm bo}\equiv r_{\rm bo}/v_{\rm sh}$, and we have used the time dependence $\Delta v_{\rm CSM}\propto E_{\rm rad}(r_{\rm sh}(t))\propto M_{\rm CSM}(<r_{\rm sh}(t)) \propto t^{3-s}$. Noting that $r\geq v_{\rm sh}t$, this term can be regarded as a first-order correction to $\Delta v_{\rm CSM}$ when $\Delta v_{\rm CSM}\ll v_{\rm sh}$. We thus neglect the contribution from the gradient of $\Delta v_{\rm grad}$, which should be a second-order correction.

As in Section \ref{sec:numerical}, we consider the velocity at the photospheric radius at $\tau=2/3$,
\begin{eqnarray}
    r_{\rm ph} \approx {\rm max}\left[R_*, \left(R_{\rm CSM}^{1-s}+\frac{2(s-1)}{3\kappa D}\right)^{1/(1-s)}\right].
\end{eqnarray}
We evaluate the maximal photospheric velocity just before the shock reaches $r_{\rm ph}$ at time $r_{\rm ph}/v_{\rm sh}$,
\begin{eqnarray}
v_{\rm ph, max}&\equiv & v_{\rm CSM,0}+\Delta v_{\rm CSM}(r=r_{\rm ph},t=r_{\rm ph}/v_{\rm sh}) \nonumber \\
&&+\Delta v_{\rm grad}(r=r_{\rm ph},t=r_{\rm ph}/v_{\rm sh}),
\end{eqnarray}
using equations (\ref{eq:Deltav_CSM_rad}), (\ref{eq:E_rad}), and (\ref{eq:Deltav_CSM_grad}). We vary $M_{\rm CSM}$ and $-t_{\rm erup}$, and fix other model parameters as in Table \ref{tab:analytical_params}. In the following sections we consider two power-law indices for the CSM profile, (i) $s=1.5$ that is expected for eruptive mass-loss like done in the numerical study, and (ii) a wind profile of $s=2$ that is often adopted in the literature.

\begin{table}[]
    \centering
    \begin{tabular}{cc}
         \hline
         Parameters & Values \\
         \hline
         CSM mass ($M_{\rm CSM}$) & $(10^{-3}-1)\ M_\odot$ \\
         CSM eruption time ($t_{\rm erup}$) & $-(1-20)$ yr \\
         CSM density profile ($s$) & [$1.5,2$] \\
         Initial CSM velocity ($v_{\rm CSM, 0}$) & $50\ {\rm km\ s^{-1}}$ \\
         Stellar radius ($R_*$) & $700\ R_\odot$ \\
         Shock velocity ($v_{\rm sh}$) & $[5000, 8000]\ {\rm km \ s^{-1}}$ \\
         Opacity ($\kappa$) & $0.34$ cm$^{2}$ g$^{-1}$ \\
         Radiation conversion efficiency ($\langle f_{\rm rad}\rangle $) & Equation (\ref{eq:f_rad_simplified})
    \end{tabular}
    \caption{Model parameters used in the analytical model. The first two parameters regarding the CSM are varied within the corresponding ranges.}
    \label{tab:analytical_params}
\end{table}

\subsection{Maximal Photospheric Velocity}
 \begin{figure*}
   \centering
    \begin{tabular}{cc}
     \begin{minipage}[t]{0.5\hsize}
    \centering
    \includegraphics[width=\linewidth]{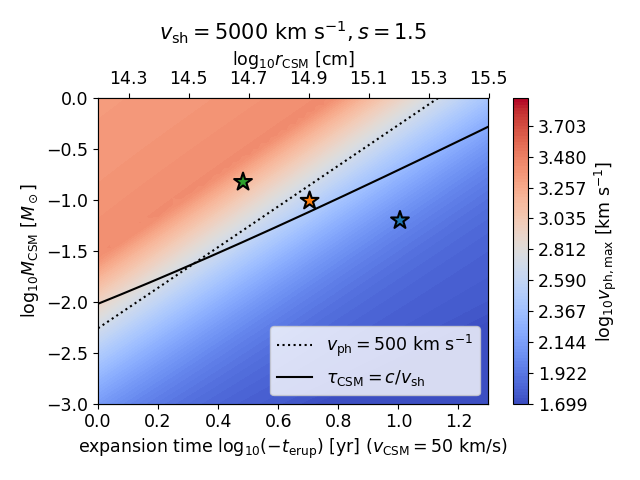}
    \end{minipage}
     \begin{minipage}[t]{0.5\hsize}
   \centering
    \includegraphics[width=\linewidth]{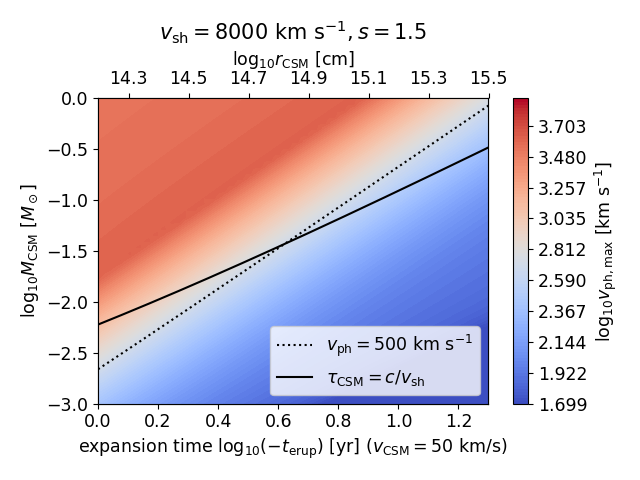}
    \end{minipage}\\
        \begin{minipage}[t]{0.5\hsize}
    \centering
    \includegraphics[width=\linewidth]{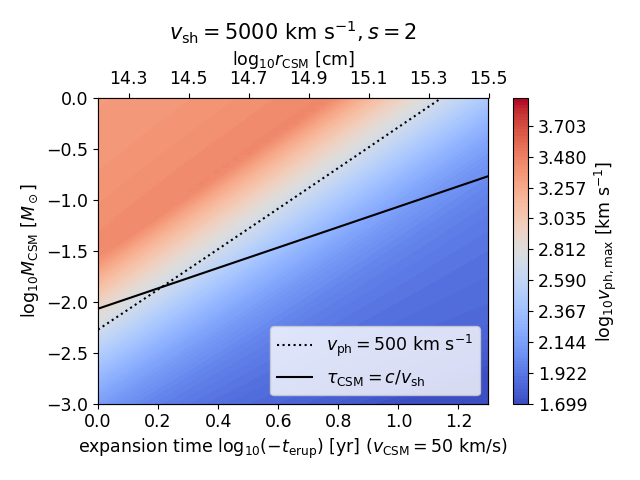}
    \end{minipage}
     \begin{minipage}[t]{0.5\hsize}
   \centering
    \includegraphics[width=\linewidth]{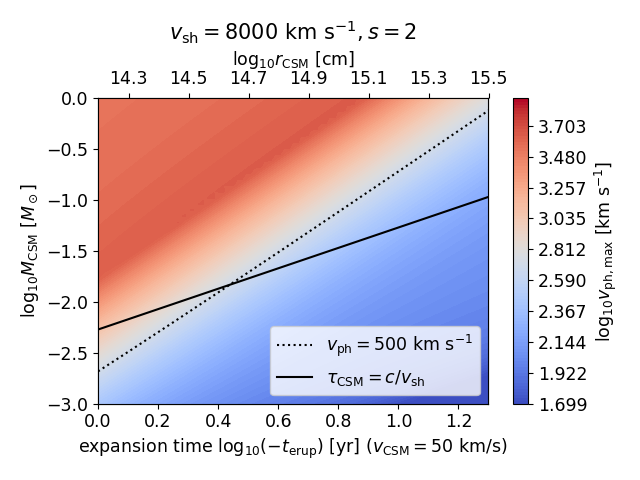}
    \end{minipage}
    \end{tabular}
\caption{Maximum velocity of the CSM at the photosphere before being swept by the shock, as a function of the CSM mass $M_{\rm CSM}$ and extent $R_{\rm CSM}$. The left and right panels show the cases for $v_{\rm sh}=5000$ km s$^{-1}$ and $8000$ km s$^{-1}$, and the top and bottom panels are for $s=1.5$ and $s=2$. The top $x$-axis is the logarithm of $R_{\rm CSM}$, and the bottom $x$-axis is the logarithm of the expansion time $R_{\rm CSM}/v_{\rm CSM,0}$ for a representative CSM velocity of $50$ km s$^{-1}$. The stars in the top left panel show the three cases of $t_{\rm erup}=-3, -5, -10$ years studied numerically in Section \ref{sec:numerical}.}
 \label{fig:v_ph}
 \end{figure*}

Figure \ref{fig:v_ph} shows $v_{\rm ph, max}$ as a function of $M_{\rm CSM}$ and $-t_{\rm erup}$ for $s=1.5$ (top panels) and $s=2$ (bottom panels). A massive and/or compact CSM generally leads to a large acceleration, which confirms the tendency seen in Section \ref{sec:numerical_acc}. 

The cases simulated in the numerical study are shown as stars in the top-left panel of the case $v_{\rm sh}=5000$ km s$^{-1}$ and $s=1.5$. These parameters are similar to the numerical study (Table \ref{table:NumericalParameters}), except for the $t_{\rm erup}=-1$ yr with a much larger $s$ that we omitted in the plot (Section \ref{sec:numerical} and Figure \ref{fig:rho_CSM}). We note that there are subtle differences in the two ways of modeling, most significant being the density profile near $R_{\rm CSM}$ which can change the location of the photosphere as well as the density slope around it. For the $t_{\rm erup}=-3$ yr ($-10$ yr) case, we find that the analytical model predicts a $\sim 20\%$ smaller (larger) $r_{\rm ph}$ than the numerical case, which leads to a larger (smaller) radiative flux. Furthermore, for the $t_{\rm erup}=-10$ yr model, there should be contribution from dissipation by mass slightly interior to $r_*$ of $\approx 9\times 10^{-3}\ M_\odot$ (15\% of $M_{\rm CSM}$), which is not included in the analytical model. Finally, our assumption of constant $v_{\rm sh}$ and $\langle f_{\rm rad}\rangle$ would also introduce some deviation from the numerical results. Despite these differences, the values of $v_{\rm ph, max}$ in the analytical model reasonably agree with the numerical results of the maximal velocity at the photosphere just before it enters the shock. Using $\langle v_{\rm sh} \rangle$ in Table \ref{table:NumericalParameters} for each of the models, the value of $v_{\rm ph, max}$ for $t_{\rm erup}=-10, -5, -3$ yr are obtained in the analytical model as $120, 440, 1700$ km s$^{-1}$ respectively. These values are to be compared with the corresponding velocities obtained in the numerical study, of $220, 540, 1300$ km s$^{-1}$ respectively.

The CSM velocity becomes saturated in the upper left corners of the plots in red, with a value comparable to the shock velocity. This corresponds to a very compact CSM with averaged mass-loss rates of $M_{\rm CSM}/(-t_{\rm erup})\gtrsim 0.1\ M_\odot\ {\rm yr}^{-1}$ for a CSM velocity of $50\ {\rm km\ s^{-1}}$. This is the case for the $t_{\rm erup}=-1$ yr model in Section \ref{sec:numerical}. In this regime, the dependence $f_{\rm rad}\propto R_{\rm CSM}^2/M_{\rm CSM}$ in our analytical model cancels out with the other factors when evaluating $\Delta v_{\rm CSM}$. This makes $\Delta v_{\rm CSM, max}$ get saturated at $\approx f_{\rm rad, 0}v_{\rm sh}/2$, which is independent of both $R_{\rm CSM}$ and $M_{\rm CSM}$.
On the other hand, acceleration is almost negligible for CSM in the lower right corners, with resultant $v_{\rm ph, max}$ less than $100$ km s$^{-1}$ for mass-loss rates of $M_{\rm CSM}/(-t_{\rm erup})\lesssim 10^{-3}\ M_\odot\ {\rm yr}^{-1}$.

Comparing between $s=1.5$ and $s=2$, the difference is dependent on the location in the parameter space. For the regions of the plot with intermediate acceleration ($v_{\rm ph, max}\lesssim 1000$ km s$^{-1}$), the predictions from the two values of $s$ are almost the same. This is because the photosphere is close to the edge of the CSM, and both $r_{\rm ph}$ and the mass interior to it are insensitive to the exact CSM profile. The largest difference is seen at the lower right region with lower density, where the $s=2$ case predicts larger acceleration. When $-t_{\rm erup}$ and $M_{\rm CSM}$ are fixed the photosphere moves more inward for $s=2$ than for $s=1.5$, resulting in larger radiative fluxes.

\subsection{Comparison with Early Spectroscopy of Interacting Supernovae}
\label{sec:result_s2}

For a wind profile of $s=2$, $D$ can be estimated by the mass-loss rate $\dot{M}$ and wind velocity $v_w$ as
\begin{eqnarray}
    D \sim 10^{16}\ {\rm g\ cm^{-1}}\left(\frac{\dot{M}}{10^{-2}\ M_\odot\ {\rm yr}^{-1}}\right)\left(\frac{v_w}{50\ {\rm km\ s}^{-1}}\right)^{-1}.
\end{eqnarray}
\cite{Boian20} conducted fitting of early optical spectra in hydrogen-rich SNe assuming $s=2$, which led them to simultaneously obtain $D$ and $v_w$.
Here we compare the observed relations of $D$ and $v_w$ with our model. As shown in the previous section, the degree of acceleration will also depend on the extent of the CSM $R_{\rm CSM}$. Here we relate this to the duration of the interaction phase\footnote{As seen in Section \ref{sec:numerical}, a compact CSM that is accelerated to a velocity comparable to $v_{\rm sh}$ can prolong the interaction phase from equation (\ref{eq:t_int}). Here we consider the samples from \cite{Boian20} where the CSM velocity is only $<1000$ km s$^{-1}$, so this complication can be avoided.}
\begin{eqnarray}
    t_{\rm CSM}&\approx&\frac{R_{\rm CSM}}{v_{\rm sh}} \nonumber \\
    &\sim& 6\ {\rm day} \left(\frac{R_{\rm CSM}}{3\times 10^{14}\ {\rm cm}}\right)\left(\frac{v_{\rm sh}}{6000\ {\rm km\ s^{-1}}}\right).
    \label{eq:t_int}
 \end{eqnarray}
Flash spectroscopy indicate that $t_{\rm CSM}$ is typically restricted to $\lesssim 10$ days for Type II SNe \citep{Khazov16,Bruch21,Bruch22}\footnote{An independent constraint of the CSM extent has been inferred from follow-up in other wavelengths, such as the non-detection of late-time radio emission for SN 2013fs \citep{Yaron17}.}, although it can be much longer for Type IIn SNe. The early spectra samples of \cite{Boian20} were taken 1--8 days after shock breakout, with a majority (12 out of 17) in the range 2--4 days. Here we consider $t_{\rm CSM}$ of $3, 5$, and $10$ days that would be realistic for these samples. Using our analytical model we calculate the photospheric velocity at day 3, for different values of $D$ and $t_{\rm CSM}$.

 \begin{figure*}
 \centering
     \begin{tabular}{cc}
     \begin{minipage}[t]{0.5\hsize}
    \centering
    \includegraphics[width=\linewidth]{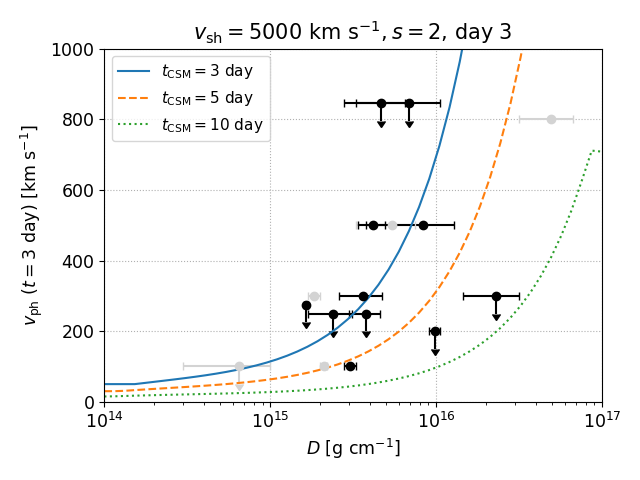}
    \end{minipage}
     \begin{minipage}[t]{0.5\hsize}
   \centering
    \includegraphics[width=\linewidth]{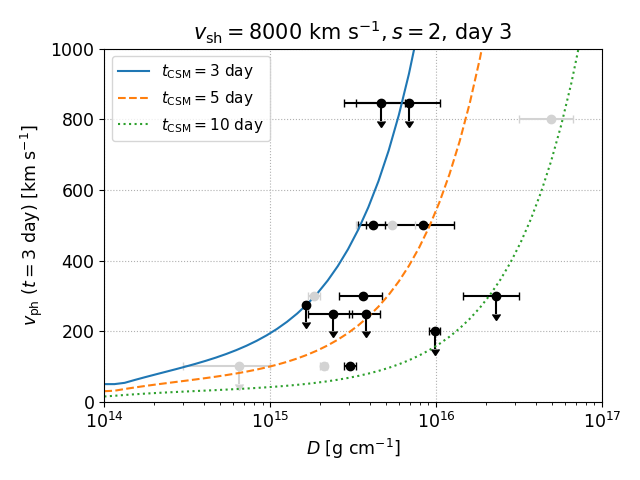}
    \end{minipage}
    \end{tabular}
\caption{Photosphere velocity $v_{\rm ph}$ at day 3 as a function of $D=\dot{M}/4\pi v_w$, calculated by the analytical model for the wind CSM case ($s=2$). Different lines show various duration of the CSM interaction phase $t_{\rm CSM}$, which is proportional to $R_{\rm CSM}$. The dots show observations and upper limits of the CSM velocity by \cite{Boian20}. The data points where the spectral observations were within day 2--4 are in black, while those outside this range are in gray.}
 \label{fig:v_ph_wind}
 \end{figure*}
 
Figure \ref{fig:v_ph_wind} shows $v_{\rm ph, max}$ as a function of the wind density parameter $D$ for $s=2$, for different values of $t_{\rm CSM}$. For low $D\lesssim R_*/\kappa$, the photosphere is close to $R_*$ and the acceleration at the photosphere is independent of $t_{\rm CSM}$. For high $D\gg R_*/\kappa$ the location of photosphere becomes close to the outer boundary of the CSM, and thus $v_{\rm ph, max}$ starts to depend on $t_{\rm CSM}$. A more compact CSM results in larger acceleration, but a shorter observable window of the upstream CSM. 

In Figure \ref{fig:v_ph_wind} we also plot the samples with spectral epoch in the range $2$--$4$ days in black, and samples outside this range in gray. Overall, given the values of $D$ inferred from spectral modelling, the CSM can be accelerated up to the observed velocities for realistic shock velocities of $5000$--$8000$ km s$^{-1}$. Radiative acceleration thus can resolve the mismatch between the observed CSM velocities (as large as $800$ km s$^{-1}$), and the much slower velocities of eruptions and winds expected from RSGs \citep{Marshall04,Mauron11,Ko22} which are believed as progenitors for a large fraction of Type II SNe \citep[][and references therein]{Smartt15}.

While a large fraction of Type II SNe likely come from RSGs, some may also originate from more compact stars like blue supergiants, in particular Type IIn SNe that constitute three of their eight samples with observed CSM velocities. A compact progenitor would of course explain the high-velocity CSM observed in some of the samples, and perhaps be a preferred explanation for ones with low $D$ that result in weak radiative acceleration. A notable example in their sample is the Type IIn SN 2010mc \citep{Ofek13}. Its CSM is found to have a velocity of $300$ km s$^{-1}$ at 8 days post explosion with a low $D=(1.7$--$2)\times 10^{15}$ g cm$^{-1}$, shown as a gray point in Figure \ref{fig:v_ph_wind}. Our model assuming a RSG progenitor is difficult to reproduce this velocity, as with such a low $D$ only a compact CSM of $t_{\rm CSM}\lesssim 3$ days can reproduce this velocity for an initial velocity of $50$ km s$^{-1}$. Thus we conclude that the CSM should already have been this fast before the SN, and plausibly originate from winds from blue supergiants (BSGs) or luminous blue variables (LBVs; \citealt{Humphreys94}). Indeed the light curves and spectra of SN 2010mc are nearly identical to SN 2009ip \citep{Smith10,Foley11,Fraser13_09ip,Pastorello13}, whose progenitor is claimed to be a BSG or an LBV \citep{Smith14,Margutti14}.

As also shown in Figure \ref{fig:v_ph}, we predict that at low $t_{\rm CSM}$ of a few days or less, the photosphere can accelerate to $>1000$ km s$^{-1}$ for dense CSM with $D\approx 10^{16}$ g cm$^{-1}$. However, the observable time window for such CSM would be shorter, making observations of such cases more difficult.

\section{Consequences for Non-thermal Emission}
\label{sec:nonthermal}

 \begin{figure*}
 \centering
     \begin{tabular}{cc}
     \begin{minipage}[t]{0.33\hsize}
   \centering
    \includegraphics[width=\linewidth]{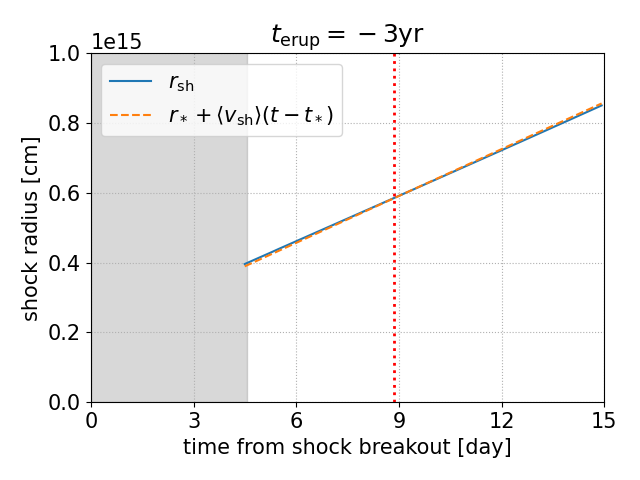}
     \end{minipage}
     \begin{minipage}[t]{0.33\hsize}
   \centering
    \includegraphics[width=\linewidth]{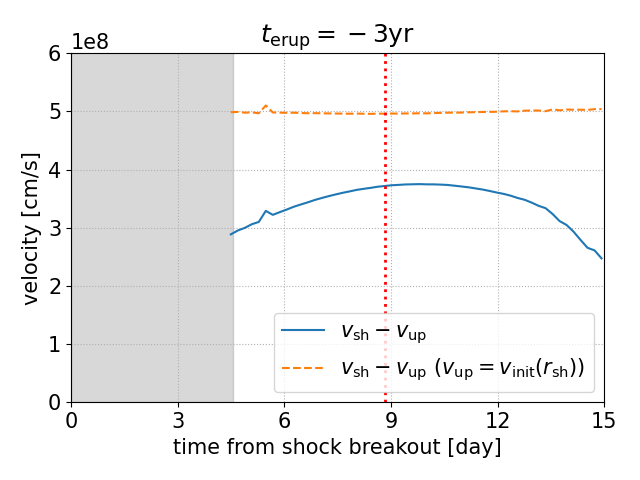}
     \end{minipage}
      \begin{minipage}[t]{0.33\hsize}
    \centering
    \includegraphics[width=\linewidth]{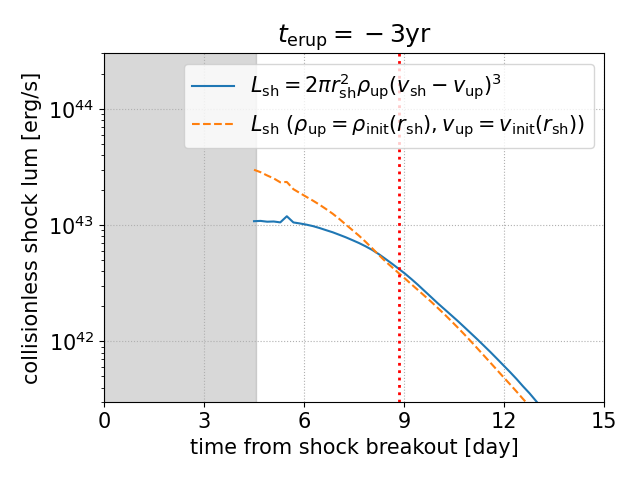}
    \end{minipage}
    \end{tabular}
\caption{Properties of the shock that forms after breakout, extracted from the numerical study with SNEC for the $t_{\rm erup}=-3$ yr model. We plot these only from $t=2t_{\rm bo}$, when we can assume that the collisionless shock is fully developed (see main text). Each panel shows the evolution of shock radius, upstream speed at shock rest frame, and shock instantaneous luminosity. For the left panel, dashed lines show uniform motion with velocity $\langle v_{\rm sh}\rangle$ with $r=R_*$ at $t=t_*$. In the middle and right panels, the dashed lines are cases assuming $v(r_{\rm up}), \rho(r_{\rm up})$ from the initial condition (i.e. neglecting acceleration), to demonstrate the effects of CSM acceleration. The vertical dotted lines show the epoch $t=t_{\rm ph}$.}
 \label{fig:collisionless}
 \end{figure*}

In this section, we use the output from the SNEC simulations to study the consequences of radiative acceleration on non-thermal emission. After shock breakout, the shock is considered to become collisionless, mediated by plasma instabilities \citep{Murase11,Katz12}. The collisionless shocks are promising sites for particle acceleration, which can give rise to non-thermal emission in the forms of multi-wavelength photons and high-energy neutrinos \citep[e.g.,][]{Murase11,Katz12,Murase14,Zirakashvili16,Petropoulou16,Murase18,Sarmah22,Kheirandish22}.

Before the shock breakout, the expanding shock is mediated by radiation due to the large optical depth of the upstream to photons \citep[e.g.,][for a review]{Levinson20}. The deceleration region is governed by the scattering of photons, with width $\delta r_{\rm sh}\sim (\kappa\rho v_{\rm sh}/c)^{-1}<r_{\rm sh}$. At $t=t_{\rm bo}$, $\delta r_{\rm sh} \sim r_{\rm sh}$ and shock breakout occurs. Afterwards the shock may not be mediated by radiation, but the formation of a strong sub-shock also requires $v_{\rm sh}>\Delta v_{\rm CSM}$, which leads to an onset of the collisionless shock at radius \citep{Murase19}
\begin{equation}
r_{\rm onset}\gtrsim \max(1,[0.5f_{\rm rad}(s-1)/(3-s)]^{s-1})r_{\rm bo}. 
\end{equation}
This implies that the collisionless shock is expected around when the breakout occurs in a rather flat density profile with $s\lesssim2$ but it is not guaranteed if $s\gg 3$. The transition would be gradual, over the timescale $\lesssim t_{\rm bo}$ that takes for radiation to fully escape from the expanding deceleration region \citep{Levinson20}. The outer part of the region loses radiation more quickly, which would likely form a subshock that is collisionless. This collisionless subshock would gradually take over as photons escape, and dominate the shock dissipation from $t\lesssim 2t_{\rm bo}$. At this regime the size of the deceleration region is of the order of the plasma skin depth $\delta r_{\rm sh}\sim 100\ {\rm cm}(\rho/10^{-13}\ {\rm g\ cm^{-3}})^{1/2}$, much shorter than resolutions that can be captured by radiation hydrodynamical simulations. In our SNEC simulations, the deceleration region is instead governed by the artificial viscosity employed between adjacent cells, resulting in its width of $\delta r_{\rm sh}\sim 10^{-3}r_{\rm sh}$.

Therefore, we can conservatively say that at least from $t\approx 2t_{\rm bo}$, the shock is largely collisionless and is capable of powering non-thermal emission by CSM interaction. We see $v_{\rm sh}>\Delta v_{\rm CSM}$, which is consistent with the analytical expectation that the shock is not mediated by radiation at $t\gtrsim t_{\rm bo}$ for $s\lesssim2$. Another consequence is that for a dense CSM with strong radiative acceleration, the shock dissipation can be somewhat weaker than for a stationary CSM, despite the latter generally being assumed in past modellings of non-thermal emission. 

To see this more quantitatively, we estimate the shock dissipation luminosity expected from the accelerated CSM at $t>2t_{\rm bo}$. Here we choose the model with $t_{\rm erup}=-3$ yr, which has significant acceleration as well as an interaction phase continuing beyond $t=2t_{\rm bo}$. From the outputs of SNEC we search for the shock radius, by first extracting the cells where the velocity difference of the fluid at $r$ and at its ``immediate upstream" $r_{\rm up}\equiv (1+\epsilon)r$ is greater than the local sound speed at $r_{\rm up}$, and then selecting the cell with the largest velocity difference. The width of the shock transition region $\epsilon$ is in reality much smaller than the resolution, but our SNEC simulations limit its width to be of order $10^{-3}r_{\rm sh}$. To fully capture both ends of the transition region with a small numerical error we adopt $\epsilon=0.01$, but after the shock becomes unmediated by radiation the resulting dissipation luminosity is largely insensitive to the choice of $\epsilon$ as long as $\epsilon\ll 1$.

Using this shock capture method at each time step of $1.7\times 10^4$ s that SNEC records the profiles, we obtain the shock radius $r_{\rm sh}$ and velocity $v_{\rm sh}$, as well as the upstream density $\rho_{\rm up}=\rho(r=r_{\rm up})$ and velocity $v_{\rm up}=v(r=r_{\rm up})$. The instantaneous luminosity by shock dissipation is then obtained by the equation $L_{\rm sh}\equiv 2\pi r_{\rm sh}^2\rho_{\rm up}(v_{\rm sh}-v_{\rm up})^3$. This luminosity sets the available budget for powering the non-thermal emission.

Figure \ref{fig:collisionless} shows the shock radius $r_{\rm sh}$, upstream speed at the shock rest frame $v_{\rm sh}-v_{\rm up}$, and the instantaneous luminosity $L_{\rm sh}$. For the latter two, in order to demonstrate the effect of radiative acceleration we also show as dashed lines the case neglecting acceleration, where we replaced $\rho_{\rm up}$ and $v_{\rm up}$ with the pre-accelerated values before core-collapse at $r=r_{\rm up}$. We find that the dissipation luminosity is suppressed at the earliest phases, with a factor of $\sim 3$ compared to the case without radiative acceleration.

One can see that the bulk acceleration slightly enhances the shock dissipation at $t>t_{\rm ph}$. 
This is because the CSM is accelerated and pushed outwards, and there is more material than the case without any feedback. This suggests that  careful investigations are necessary to understand the formation of collisionless shocks at $r_{\rm bo}>R_{\rm CSM}$. Naively, when the breakout occurs at $r_{\rm bo}>R_{\rm CSM}$, the density profile is so steep (i.e., $s\gg 3$ in our case) that the dissipation via collisionless shocks would essentially be negligible in the sense that $r_{\rm onset}\gg r_{\rm bo}$. However, the net effect is unclear due to feedback, and more detailed simulations with better resolutions capturing sub-shocks are necessary.

\section{Conclusion}
\label{sec:conclusion}
In this work we considered radiative acceleration of CSM in SNe powered by circumstellar interaction. With a numerical approach we presented the results of the radiation hydrodynamical simulation, SNEC. We then constructed an analytical model, calibrated by the numerical simulations, which enables us to systematically study the parameter space and compare the results with the observed CSM velocities from early spectra of interacting SNe. From these modelling we obtained two key results, one on the observed CSM velocities in interacting SNe and the other on the possible suppression of the non-thermal emission from these signals.

From both numerical and analytical models, we found that the radiative acceleration is generally stronger for a compact and/or massive CSM. The range of the parameter space in the mass and extent in the CSM helps us explain the broad range of the CSM velocity ($\lesssim 100$ to $\sim 10^3$ km s$^{-1}$) observed in interacting SNe.

For an even denser CSM with mass loss rates of $\dot{M}\gtrsim 0.1\ M_\odot\ {\rm yr}^{-1} (v_{\rm CSM}/100\ {\rm km \ s^{-1}})$, this acceleration may significantly suppress the dissipation through collisionless shocks that are expected to form soon after shock breakout in the CSM. The results would be useful for better modelling particle acceleration and multi-messenger signatures of these shocks, which would be a subject of forthcoming studies.

For our mechanism a larger radiative flux results in larger acceleration, so the correlation between the line width of the CSM and the radiated energy (or peak luminosity) may be expected from analyzing a large sample of interacting SNe. One uncertainty is the extent of the CSM being different for each SNe, which can alter the photospheric radius (i.e. radiative flux) and possibly smear the possible correlation. Nevertheless, if such correlations are observed this would be strong evidence for radiative acceleration being responsible for the observed CSM velocity.

As demonstrated in the end of Section \ref{sec:result_s2}, we may be able to constrain the initial velocity of the CSM if we can infer both its velocity and density (alternatively, mass-loss rate if wind profile is assumed). This can give independent clues about the progenitor, especially for Type IIn SNe which are under debate and possibly diverse \citep[for reviews see][]{Smith14_review, Smith17}.

\begin{acknowledgments}
The authors are deeply indebted to Viktoriya Morozova for early contributions, her supports on running the SNEC code, and many valuable discussions. The authors also thank the anonymous referee for the insightful comments. D.T. thanks the hospitality at the Pennsylvania State University and YITP at Kyoto University, where this work was mainly done. This work was supported by the JSPS Overseas Challenge Program for Young Researchers, JSPS KAKENHI grant Nos. JP19J21578, MEXT, Japan, and the Sherman Fairchild Postdoctoral Fellowship at Caltech (D.T.). 
The work was partly supported by the NSF Grants No.~AST-1908689, No.~AST-2108466 and No.~AST-2108467, and KAKENHI No.~20H01901 and No.~20H05852 (K.M.)
\end{acknowledgments}

\bibliography{references}
\bibliographystyle{aasjournal}

\end{document}